# Surface plasmon resonance sensor based on a novel prism for the detection of a broad range of polymers

Natalia A. Gutierrez Andrade, Yunfeng Nie, Wendy Meulebroeck, Heidi Ottevaere*

Brussels Photonics (B-PHOT), Vrije Universiteit Brussel and Flanders Make, Dept. of Applied Physics and Photonics, Pleinlaan 2, 1050 Brussels, Belgium.

*Corresponding authors: Heidi.Ottevaere@vub.be, Natalia.Andrea.Gutierrez.Andrade@vub.be

**Abstract**

Microplastics are pervasive pollutants that pose threats to the environment and human health. Detecting them reliably represents a crucial step toward finding solutions. While traditional techniques like vibrational and mass spectroscopies are time-consuming and have limitations in analysing polymers with pigment additives, surface plasmon resonance (SPR) shows promise due to its high sensitivity and rapid detection capabilities. However, existing SPR methods rely on one output variable and fail to determine various microplastic parameters simultaneously. To address this uncertainty between size and polymer type, we propose a prism design that enables SPR resonance in a refractive index range of 1.402–1.56 RIU (Refractive Index Units) for polymer detection. Experimental verification of the prism's dynamic range using refractive index oils presented an agreement of 98.47% with theoretical analysis. Furthermore, we acquired SPR curves from ten different polymer types, including Nitrile, Viton and Tyre rubber. These results demonstrate SPR's potential to identify polymers and its ability to detect even pigmented materials—often challenging for other techniques. The imaging capabilities of the novel prism are presented, indicating its potential for future imaging of microplastics. Our approach holds promise for advancing microplastics analysis by enabling comprehensive assessment of particle parameters, such as polymer type, size, quantity and shape, and expanding the range of detectable polymer types.

1. ## Introduction

Microplastics consist of small synthetic polymer particles released into the environment. Their size allows them to disperse across diverse ecosystems [1–3], exposing organisms [4,5] and humans to their hazardous effects [6–8]. Prior studies have indicated that their toxicity is related to various parameters such as size, shape, polymer type, and quantity [9–11]. Therefore, all these parameters should be addressed to enable comprehensive toxicological analysis. To analyse microplastics, these must first be isolated from the sample matrix and subsequently characterised [12]. Both stages, isolation and analysis, are complex procedures that depend on many variables. Consequently, numerous analytical methods have been proposed to address all microplastics parameters [13–15]. However, existing methods still face limitations, which makes it difficult to define a standard analysis method [9,13,14,16] and further efforts are necessary to achieve a suitable analytical approach.

Micro Raman spectroscopy (μ-Raman) [14,17] and micro-Fourier transform infrared spectroscopy (μ-FTIR) [18,19] are commonly used for microplastic analysis. These techniques provide spectral information for polymer identification and microscopic images, enabling particle enumeration and characterization. However, they are time-consuming due to the required scanning and complex data processing, resulting in analysis durations exceeding 10 hours per sample [20,21]. Additionally, their spatial resolution limits affect detectable particle sizes. Furthermore, pigment additives in the microplastic samples can alter spectral signals, leading to underestimation [22]. To address this underestimation, mass spectroscopy (MS) techniques measure the polymer's mass content per sample

[23–25]. However, this approach has limitations—some polymers lack unique degradation products for specific identification, and environmental samples may yield false positives due to the presence of mixed materials [26,27]. Moreover, MS methods do not efficiently provide morphological characteristics of the particles. Therefore, new methods are being studied to overcome these challenges.

Surface plasmon resonance (SPR) is being explored for microplastics analysis due to its high sensitivity and rapid detection capabilities [28]. Initial studies demonstrate that SPR signal shifts, combined with mathematical models or chromatography separation, offer sufficient sensitivity for characterising micro- and nano-plastics [29–31], even at real samples' concentrations [32]. Additionally, SPR imaging enables quantification, particle shape and size analysis [33,34]. These advantages make SPR a promising technique for microplastic analysis, potentially overcoming challenges related to particle size and reducing analysis time. However, current results are limited because they rely on one output signal to address all microparticle characteristics and thus they can't be differentiated. For instance, Zybin et al. report that the intensity signal shift correlates with the polymer type for particles of the same size [35]. In contrast, Maley et al. [33] demonstrate that the same signal can be associated with particle size for particles made of the same material. Consequently, the technique struggles to analyse sets of particles with varying sizes and polymers, which is the case of real microplastics samples.

Further research is essential to optimize SPR performance for microplastic analysis. Previous studies have been limited by a bio-detection approach, where resonance was achieved with the sample matrix rather than directly with the particles. Traditional SPR biosensors target resonance with water-based buffer solutions (around 1.33 RIU), where the sample of interest is dissolved [36]. However, SPR fundamentally relies on detecting materials' dielectric constants, which correspond to their refractive index (RI) [37]. Peiponen et al. [38] suggest that the refractive index of polymers, and thus their dielectric constant, could be a key identification parameter. This represents a clear distinction from biosensing approaches, as polymers typically exhibit higher refractive indices (approximately 1.49 to 1.58 RIU at 635 nm). Notably, to the best of our knowledge, SPR has not yet been used for direct polymer detection. Expanding SPR's RI range using a high refractive index prism, as demonstrated by Canning et al. [39], could allow for the distinct acquisition of SPR signals for each polymer type, thereby enabling a comprehensive analysis of microplastic parameters. Therefore, polymer detection is a crucial step toward enhancing SPR's capabilities for effective microplastic analysis.

In this paper, we demonstrate SPR resonance with polymers. Initially, we analysed the Kretschman configuration model to define the material and geometry of a prism suitable for polymer detection through angular interrogation. The prism design undergoes initial evaluation via simulations. Subsequently, we test the prism's performance using refractive index oils (RIOs) and compare the experimental SPR curves to the simulated results. Our results confirm that the broad dynamic range of the prism encompasses the refractive index range of various polymers. Additionally, we present the SPR curves for ten polymer types, including Polymethyl methacrylate (PMMA), polypropylene (PP), polyvinyl chloride (PVC), low-density polyethylene (LDPE), polyurethane (PU), Viton, Nitrile, and Tire rubber (TR). Finally, we explore the potential combination of our designed prism with SPR imaging, which may lead to a comprehensive analysis of microplastics parameters. Here we demonstrate that the novel prism developed would enable the detection of various polymer types through their different resonance angles. We expect that these results could be used to further analyse the dielectric constant of polymers as an identification parameter even for the cases of non-transparent samples. Ultimately, these results may enhance the SPR approach to microplastics detection and enable further benefit from its advantages, such as short analysis time and high sensitivity, for microplastic analysis.

## 2. Materials and Methods

### 2.1 Materials

To verify the performance of our system, we conducted initial measurements using refractive index oils (RIOs) from Cargille. Specifically, we tested various refractive index oils (RIOs) with refractive indices of 1.44, 1.456, and 1.458 from series AA, as well as those with refractive indices of 1.46, 1.47, and 1.49 from series A, and 1.55 and 1.56 from series E.

In our study, we selected various polymer types based on the minimum polymer data set recommendation by Schymanski et al. [14]. Pure PP was acquired from Sigma Aldrich. LDPE pure resin LD150 was obtained from Exxon Mobil. Other samples were derived from commercial items including PMMA, COC, PVC, PU, TR, Viton, and Nitrile. Their composition was ensured through ATR- FTIR spectra analysis. TR, Viton and Nitrile were chosen due to their challenging analysis using other spectroscopic methods [40].

Polymer measurements contrast with traditional measurements in SPR which are made on liquids [41]. To measure the polymer samples, we needed to ensure a surface of contact with the SPR chip large enough to fit the laser spot. Therefore, we proposed different methods to achieve sample-sensor contact to perform measurements. The sample was pressed against the sensor's gold surface in the case of PMMA, PU, TR, Viton, and Nitrile. For LDPE and PP, the sample was placed on the SPR chip and melted in an oven at 160°C and 180°C respectively, later the oven was let to cool down with the sample inside. Other polymers like PS and PVC were dissolved in toluene in a ratio of 100mg/ml. The dissolved polymer was subsequently placed on the SPR chip and dried under a fume hood at room temperature. We measured the FTIR spectra, in the possible cases, before and after these procedures to ensure that the processes applied did not alter the polymers.

### 2.2 Simulation procedure

To collect refractive index (RI) data for our simulations, we measured the refractive indices of both refractive index oils (RIOs) and polymers. The measurements were conducted using an Anton Paar Abbemat Multiwavelength Refractometer, at 20°C, for the wavelengths 486.8nm, 532.1nm, 589.3nm, and 655.7nm, with four repetitions for each wavelength. The obtained values were then fitted using the Conrady equation within the Ansys OpticStudio Glass fitting tool [42]. Subsequently, we calculated the refractive index at 635 nm. The dielectric constant values of PVC and PS were sourced from the literature [43,44]. Additionally, the dielectric constant used for the gold coating was reported by a previous study [45], which specifically applies to thin layers and is compatible with the ones used in our experiments. These values are summarized in Table 1. Later these values would serve as a reference for the experimental results achieved as well.

*Table 1. Refractive index (RI) and dielectric constant (DC) values at 635nm were used in the simulations and design of the sensor.*

| Refractive index oils (RIO) | | | | | |
|---|---|---|---|---|---|
| Reference | RI [RIU] | DC | Reference | RI [RIU] | DC |
| RIO 1.402 | 1,3984 | 1,9555 | RIO 1.47 | 1,4706 | 2,1627 |
| RIO 1.44 | 1,4401 | 2,0739 | RIO 1.49 | 1,4871 | 2,2115 |
| RIO 1.456 | 1,4563 | 2,1208 | RIO 1.55 | 1,5488 | 2,3988 |
| RIO 1.458 | 1,4583 | 2,1266 | RIO 1.56 | 1,5579 | 2,4271 |
| RIO 1.46 | 1,4603 | 2,1325 | | | |
| **Polymers** | | | | | |

| Type | RI [RIU] | DC |
|---|---|---|
| PMMA | 1,4903 | 2,2210 |
| COC | 1,5315 | 2,3455 |
| PVC [43] | 1,54 | 2,3716 |
| PET | 1,5723 | 2,4721 |
| PS [44] | 1,5874 | 2,5198 |
| Metal | | |
| Gold [45] | NA | -12,797+i1.398 |
| Titanium [46] | NA | -6.887+i20.43 |
| Glass | | |
| Schott SF11 | 1.7783 | 3.1623 |

We used the open-source software Winspall from ResTec [47] to simulate the surface plasmon resonance (SPR) curves for both the refractive index (RI) oils and the polymer samples. This software is based on transfer matrix methods and Fresnel coefficients. The simulation parameters were configured for a laser wavelength of 635 nm, spanning an angular range from 35° to 90° to cover the desired RI range. Each curve was sampled at a resolution of 400 points. The prism specifications involved a half-cylinder prism or a triangular prism with an apex angle of 20°, corresponding to the designed 80° input surface angle.

## 2.3 Optical measurement setup

The optical system comprised an illumination segment, followed by the surface plasmon resonance (SPR) sensor, and finally, the detector, as depicted in Figure 1. The collimated beam at 635 nm (Thorlabs, LP635 SF8) was polarized using a linear polarizer (Thorlabs, LPVIS050-MP2). The polarization angle was controlled by a liquid crystal polarization rotator (LCPR) (Thorlabs, LCR1633). The beam then entered the designed prism, which was attached to a commercial bare gold SPR chip from Xantec reference SCR-AU-HRI. The SPR chip consists of a 300μm SF10 glass substrate, a 1-2nm adhesion layer of Ti and a 44nm layer of Gold. This was the substrate was the closest available to SF11 glass. We used the 1.80 refractive index oil from Cargille's series M to bind the SPR chip to the prism. The reflected beam intensity was measured using a photodiode detector (Thorlabs, S121C).

The right polarization

To enable angle variation of the sensor, we secured the prism to a custom-made holder, which was mounted on a mechanical rotation stage (Artisan reference URM80PE). To align the polymer pieces and achieve the wide field of view frames, we increased the beam diameter using a 10x microscope objective and filtered it through a 15 μm pinhole from Newport. The images were acquired with a Ximea XiD MD120MU-SY CCD camera.

The wavelength of 635nm was chosen because, at shorter wavelengths, the dielectric constant of polymers is larger, implying that the sensor can also operate at longer wavelengths (NIR) if needed. Additionally, shorter wavelengths yield a shorter propagation length, which results in a higher resolution in SPR imaging [48,49]. Although larger wavelengths are typically preferred due to enhanced full-width half maximum (FWHM) and sensitivity of the SPR curves [50], these wavelengths come with the trade-off of increased propagation length. By considering shorter wavelengths (visible red ~630nm) for the prism design, the prism becomes versatile for a wider wavelength range, accommodating both imaging and single-point measurements as needed.

## 2.4 Data analysis of experimental readings

Each curve presented is the average of five subsequent repetitions. For each point in a repetition curve, we normalized the intensity signals between S (transverse electric) and P (transverse magnetic) polarizations. Afterwards, we applied a polynomial curve fitting using MATLAB curve fitting toolbox near the minimum point region employing a fifth-order polynomial function. The goodness of fit was assessed by ensuring an R-value above 0.99 for all cases. From these fitted polynomials, we calculated the minimum point of each repetition curve. The angle associated with the minimum intensity value from each of the five repetitions was then averaged to determine the resonance angle, along with the standard deviation (σ), for each sample.

## 3. Results and discussion

### 3.1 SPR prism design

SPR relies on the excitation of surface plasmon polariton waves (SPPWs). These waves occur at the interface between a metal and a dielectric, combining plasmons within the metal and polaritons inside the dielectric. Consequently, SPPWs depend on the dielectric constants of both materials. One configuration commonly used to generate SPPWs for sensing is the Kretschmann configuration. It involves a prism coated on one of its surfaces with a thin metallic layer, as depicted in Figure 1. The thin metal layer serves as the metal part of the interface where SPPWs occur. When monochromatic light (P-polarized with respect to the YZ plane) illuminates one of the prism's uncoated faces, the beam reflects off the metal surface, and part of its electric field, known as the evanescent field, is transferred into the metal layer. If the resonance condition described in equation 1 is satisfied, a SPPW is generated. When an SPPW takes place, a shift of intensity or phase will be observed in the reflected beam by a detector. The intensity shift of the reflected beam is used as the sensor output.

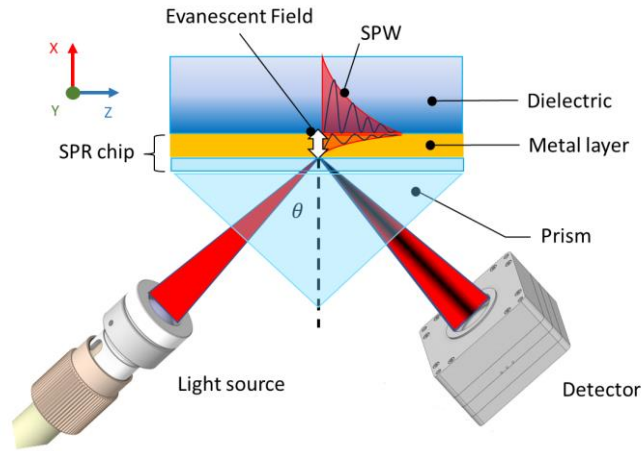

*Figure 1. Kretschman configuration setup depicting how the evanescent field transfers to the metal coating and generates resonance of SPPW.*

$$\frac{\omega}{c} n_p(\omega) \sin(\theta) = \frac{\omega}{c} \sqrt{\frac{\varepsilon_m(\omega) \cdot \varepsilon_d(\omega)}{\varepsilon_m(\omega) + \varepsilon_d(\omega)}} \qquad (1)$$

The left side of equation 1 is the light wave vector component parallel to the interface (z-axis) in terms of the angle of incidence (θ), the refractive index of the prism ($n_p$), the light angular frequency ω, and the speed of light in vacuum c. The right side of the equation is the SPPW wave vector which depends on the real part of the metal's dielectric constant $\varepsilon_m$ and the dielectric constant of the sample

$\varepsilon_d$, which is the square root of the refractive index of the sample $n_s$. The prism refractive index ($n_p$) increases light's wave vector, enabling it to equal the wave vector of the surface plasmon polariton wave (SPPW) fulfilling the resonance condition [51].

Equation 1 illustrates how the wave vector component in z is modulated by the incidence angle θ to match the momentum of the SPPW. For a given sensor, the maximum wave vector it can provide is reached when the light is parallel to the z-axis and is limited by $n_p$ at the specific wavelength used. To achieve resonance for a specific sample, one can calculate the SPPW momentum using an approximate value of $\varepsilon_m$. Afterwards, a resonance angle can be assigned to the calculated SPPW momentum. Finally, by replacing these values in equation 1, one can calculate the required value of $n_p$ to achieve the required resonance.

To further visualize this analysis, Figure 2 presents how light momentum is modulated by $n_p$ for various materials at 635nm and compares it to the SPPW momentum of PMMA and PS. It depicts how for BK7 with n=1.55, light does not reach the wave vector value required for SPR resonance with PMMA. However, for materials with higher RI, it would be possible to achieve SPR resonance not only with PMMA but also with all the range of polymers' RI up to PS. This is the case of materials like SF11, SF66, or ZnSe. Therefore, the selection of the prism material is a relevant parameter to enable polymer detection.

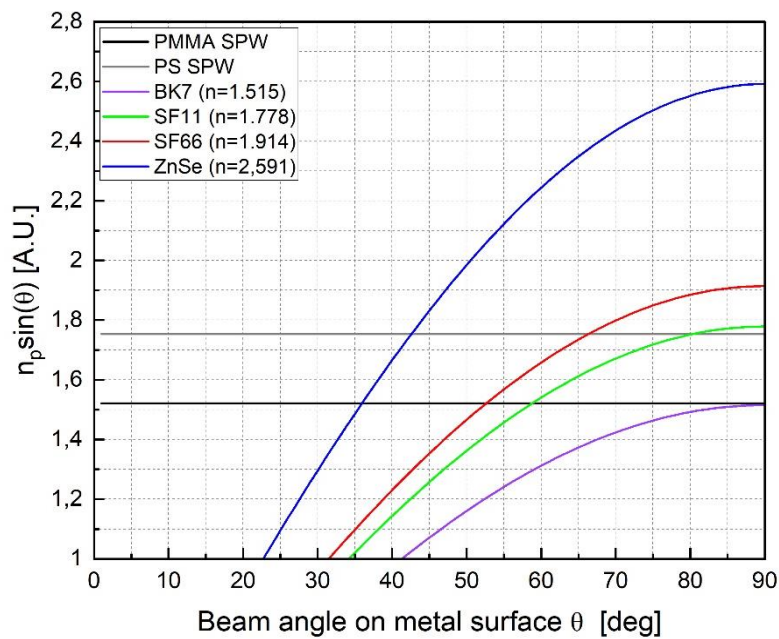

*Figure 2. Light momentum modulated by the refractive index (n) of various glasses at 635nm and the incidence angle of the beam on the sensor surface compared with the SPPW wave vector of PMMA and PS.*

To design the prism for polymer detection, we listed a range of dielectric constants from different polymer types, as depicted in Table 1. Subsequently, we selected the highest value from this range, which corresponds to Polystyrene (PS) with an RI of 1.5874. The high RI of PS was chosen to ensure that the sensor could detect polymers with lower refractive index. Provided that PS has the highest RI, of the polymers found in literature, then it would also have the highest wave vector for SPR resonance. Therefore, if the sensor can reach this high wave vector it can also achieve the wave vector of other polymers by decreasing the incidence angle θ. Next, we calculated the Surface Plasmon Polariton momentum of PS, considering the dielectric constant of gold ($\varepsilon_m$=-13.032 at 635 nm). We then obtained the value of $n_p$ by considering an 80° resonance angle, following literature recommendations to achieve a quality SPR signal [39] and using the calculated SPPW momentum. Finally, we chose Schott glass NSF11 as the prism material due to its close refractive index to the calculated n value.

After selecting the prism material, we performed initial simulations with Winspall software to calculate the resonance angles of polymers for a half-cylinder geometry. This geometry was chosen because the beam is normal to the surface at any angle, therefore, the beam is not refracted into a different angle than the incidence one. Consequently, the calculated angles indicate the angle range that should be aimed with the prism geometry of the prism designed. The obtained SPR curves are shown in Figure 3. From this simulation, it was possible to notice the angle range targeted with the sensor between 65° to 80°.

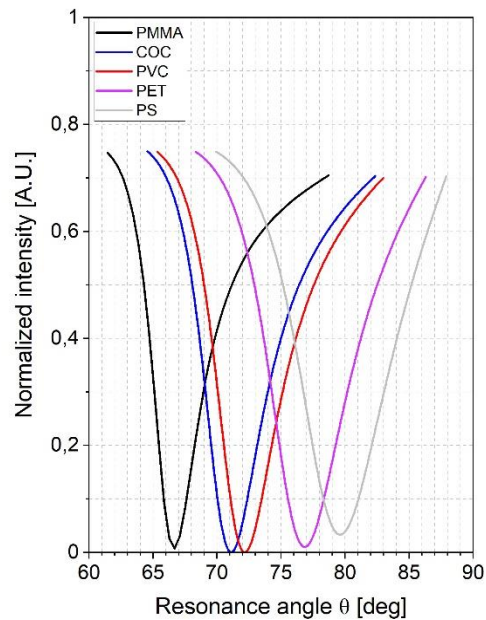

*Figure 3 Simulated SPR curves of polymers for an SF11 half-cylinder prism.*

The final aspect of the prism considered is the geometry. In order to achieve a system capable of SPR imaging a flat-face prism is required, which generally is a triangular prism. A system capable of imaging is of interest for microplastics detection to provide the particles' size, shape and quantity. In their work, L. Laplatine et al [52] demonstrate that the output surface of the prism should be perpendicular to the reflected beam to perform SPR imaging with high resolution. Therefore, they suggest the use of an isosceles prism, which continues to be used [34,52,53]. This geometry is focused solely on the output surface and overlooks its impact on the sensor's dynamic range. However, the effect of the input surface becomes relevant for a high dynamic range application. For this reason, it is necessary to analyse the geometry of the prism and its impact on the resonance angles detected.

Figure 4 shows how for flat surface prisms, the external beam incidence angle α is not the same as the angle θ of the beam incident on the coated surface due to refraction on the input surface. Thus, if the external beam is incident on 0°, or parallel to the x-axis, it will be refracted to a larger angle on the coated surface. On the other hand, if the beam enters the prism parallel to the z-axis it will be refracted into a smaller angle. In consequence, the angle θ will not vary in the same range and there would be a limited range of values of θ to achieve resonance. Due to Snell's law, the angle of the input surface γ alongside $n_p$ constrain the angle range of θ in which SPR measurements can be done. As $n_p$ was chosen to ensure the resonance with various polymers, then γ should be chosen to enable the refracted beam incidence on the desired resonance angle range (Figure 2). By considering the Snell law and the input surface angle γ, we deduced equation 2 which relates the mentioned angles α, θ, and γ.

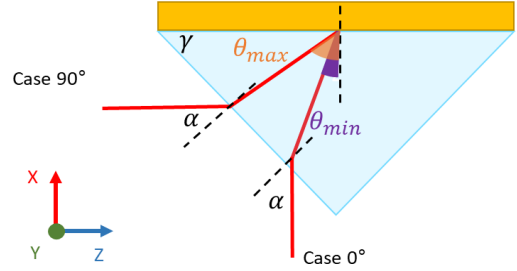

*Figure 4. Refraction of the light beam on the input surface narrows the angle range of θ that can be used to achieve SPR resonance with the sample.*

$$\theta = \gamma - \sin^{-1}\left(\frac{\sin(\gamma - \alpha)}{n_p}\right) \quad \alpha < \gamma \quad \theta = \gamma + \sin^{-1}\left(\frac{\sin(\alpha - \gamma)}{n_p}\right) \quad \alpha > \gamma \quad (2)$$

Figure 5 illustrates the relationship between the angle range of θ and γ as described in equation (2). The figure depicts that γ restricts the angle range of θ, with an increase in γ leading to higher values of θ's range. For example, when γ = 45°, it is not feasible to achieve resonance at angles greater than θ = 70°. However, higher resonance angles θ become possible when γ ≥ 70°. It's important to note that setting γ to 70° would also raise the minimum measurable angle of θ. Therefore, γ should be selected based on the refractive index (RI) range targeted by the sensor. Since resonance angles between 65° to 80° are necessary to achieve resonance within the RI range of polymers shown in Figure 3, γ was set to 80°.

The output surface angle of the prism was the average of the resonance angles of polymers to reduce the distortion for the imaging tests, the angle calculated was 70°. The geometry of the prism design is depicted in Figure 6. The prism was later custom manufactured by the company UltiTech Sapphire Corporation in China.

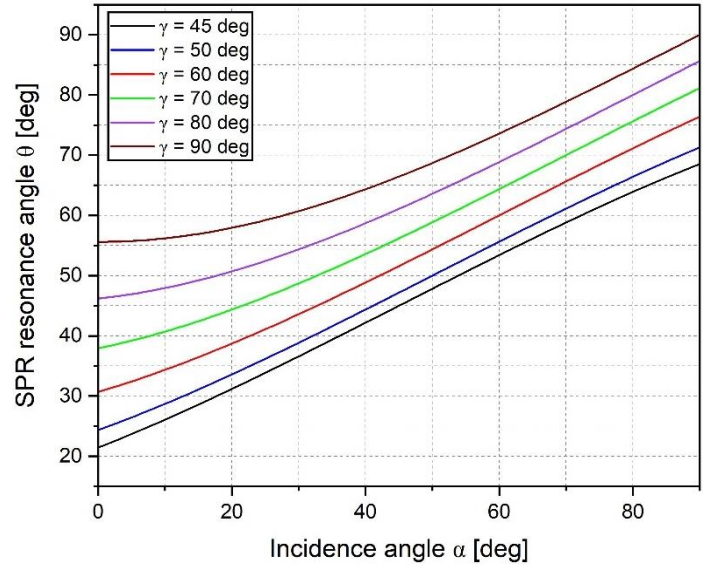

*Figure 5 Relation between the refracted angle (θ) due to the incidence angle (α) for various surface input angles (γ).*

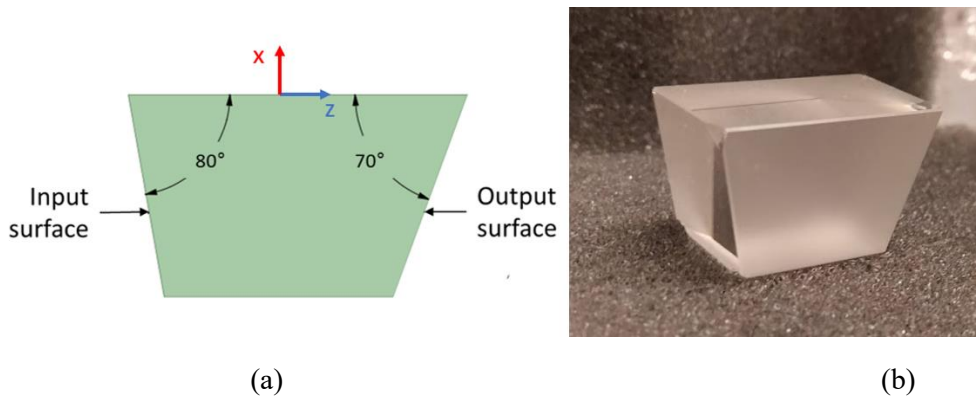

(a)                  (b)

*Figure 6 SPR Prism to achieve SPR resonance curves of polymers (a) design and (b) manufactured piece.*

An initial test of the design was performed through simulations. The dielectric constant parameters used correspond to the ones in Table 1, described in the method section. The geometry of the prism was specified as a triangular prism with an angle of 20°. 20° corresponds to the apex angle in an isosceles triangle geometry where the other angles have 80°, as the input surface angle selected. The results are depicted in Figure 7. The curves achieved demonstrate the feasibility of the design to perform the detection of polymers with SPR.

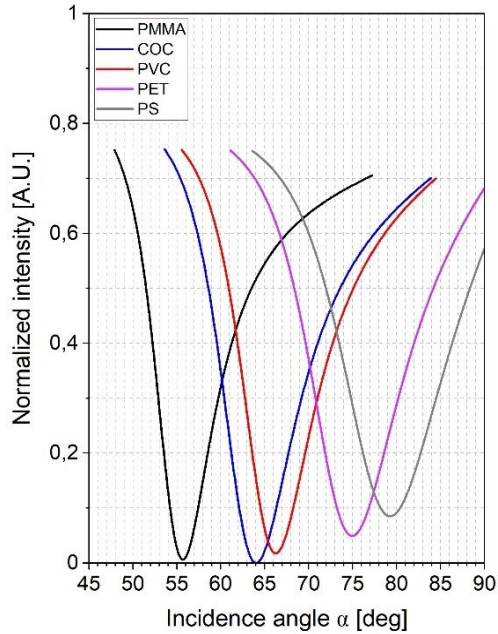

*Figure 7 Simulated SPR curves of various polymers for the prism designed in SF11 Schott glass.*

### 3.2 Sensor characterization with RIOs

The initial tests of the sensor were conducted to assess its dynamic range, which aimed to detect RI between 1.49 to 1.58 RIU and to determine how the design influenced the sensitivity and LOD. These measurements were performed with two sets of RIOs from Cargille. The first set aimed at a wide RI range (1.402-1.56 RIU) to assess the sensor's dynamic range. The second set consisted of a narrow RI range (1.44-1.49 RIU) to identify the sensitivity and LOD. For the measurements, the prism was attached to a fluidic system and each RIO curve was measured five consecutive times. Afterwards, the fluid cavity was cleansed with isopropyl alcohol before introducing the next RIO sample into the sensor. The curves acquired for each set are depicted in Figures 8 and 9.

Aside from the experimental measurements, simulations of the SPR curves for the various RIOs were made as reference points for the experimental results. The curves simulated are depicted in Fig. 8a and 9a. The minimum resonance angles calculated are presented in Figs 8c and 9c.

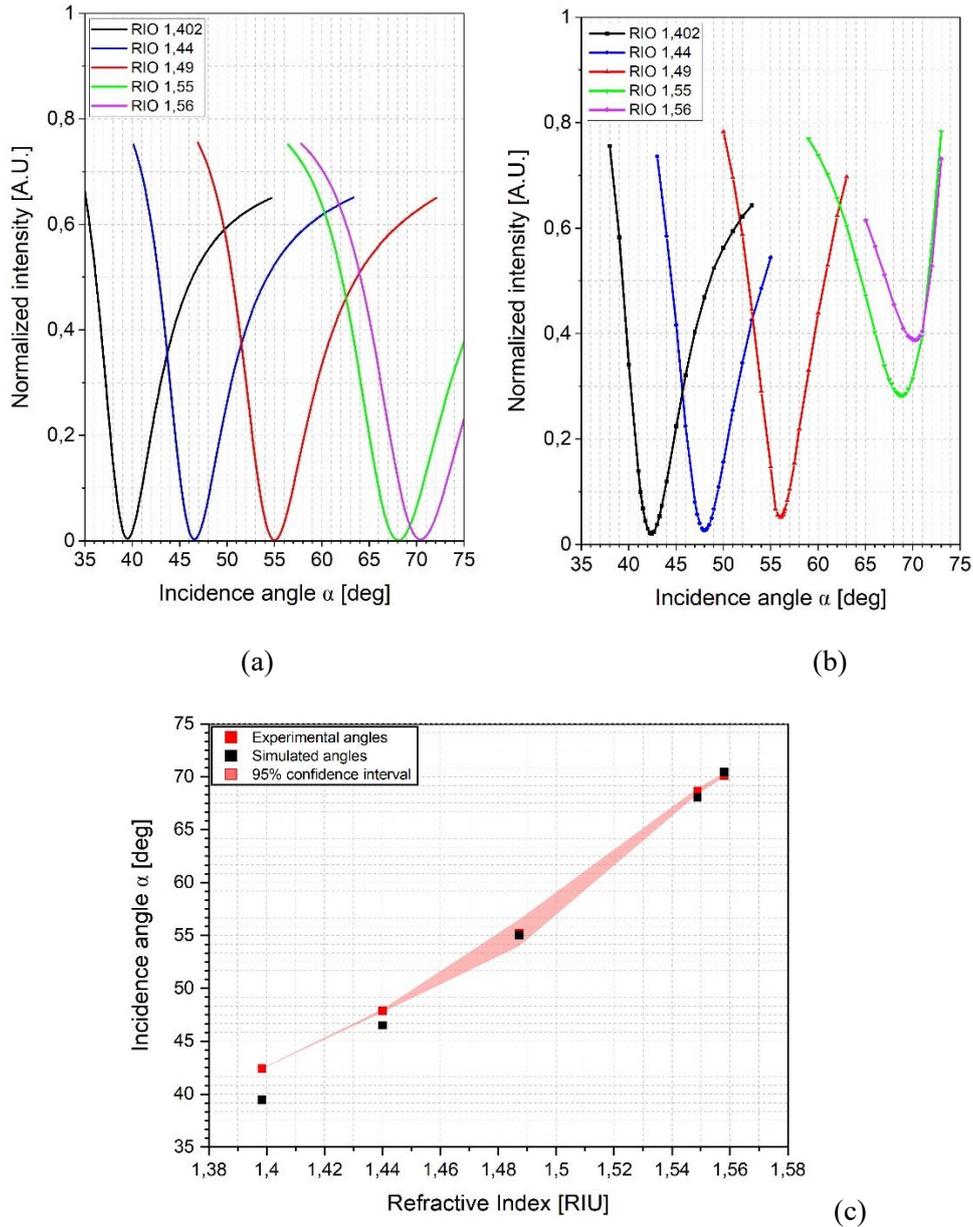

*Figure 8 Simulated (a) and experimental (b) SPR curves of RIOs from 1.402 to 1.56 RIU to assess the dynamic range of the sensor. (c) Resonance angle comparison between simulation and experimental results.*

Figure 8 depicts the results achieved on both simulation and experimental measurements for the wide range set of RI. In general, the results confirm that the defined design can perform measurements in the RI range chosen. The measurements were performed between 1.402 and 1.56 RIU, which is a range that encompasses the RI of various polymers depicted in Table 1. The agreement between the simulations and the experimental results validates the method used for the design to achieve resonance at the RI range of polymers. A difference of a few degrees is observed between experimental and simulated angles, the largest difference was 2.93° for RIO 1.402. This might be due to the difference between the dielectric constant of gold used in the simulation and the experiment, which was also noted by Zybin et al [35]. It was observed that the curves of higher RIO 1.55 and 1.56 have a resonance intensity minimum higher than the ones in the simulation. This could be related to the interference effects from multiple layers of the sensor prism and SPR chip, potentially overlapping the measured data. The interference effect is particularly observed at larger resonance angles where the difference in refractive index between the layers generates total internal reflection.

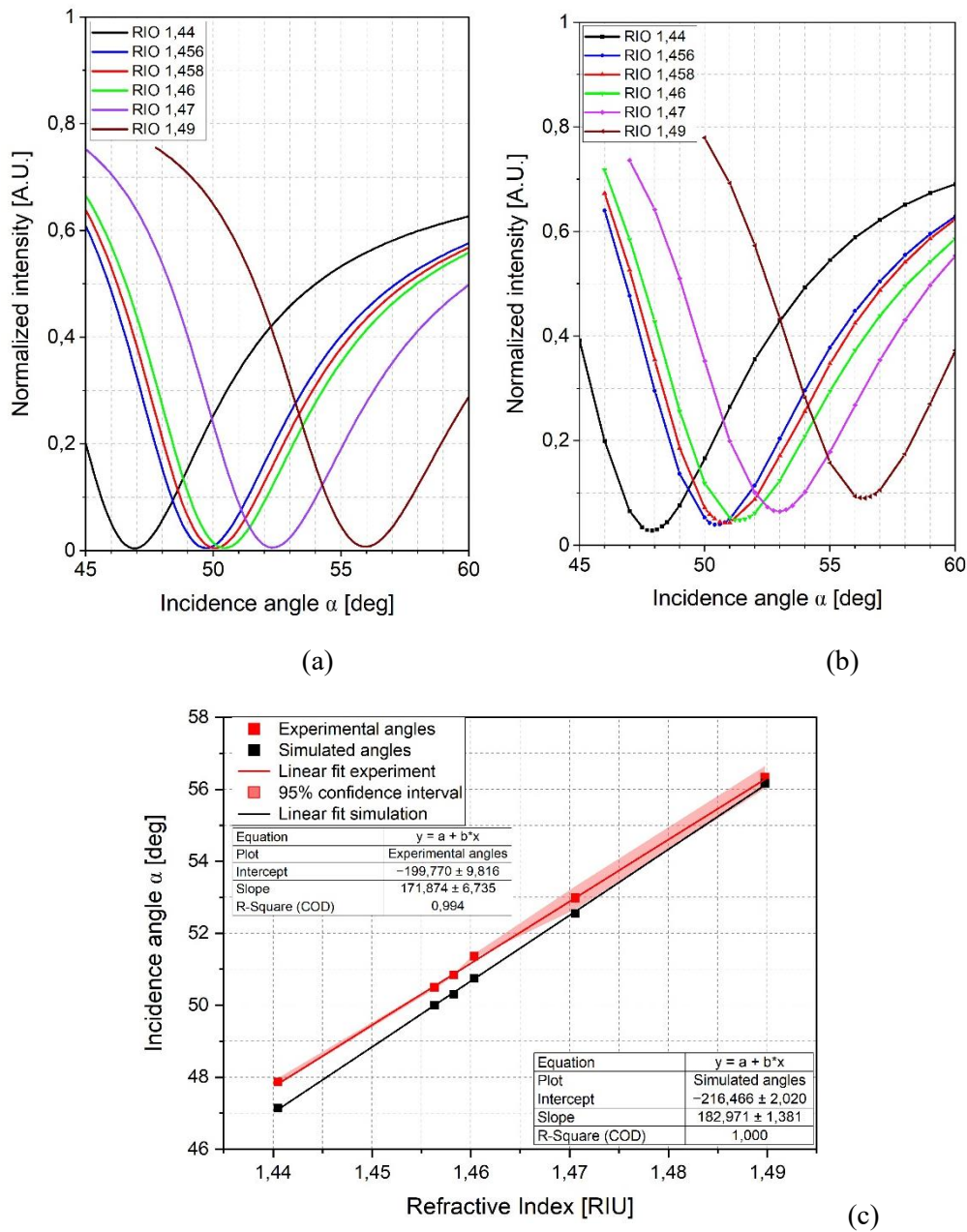

Figure 9 Simulated (a) and experimental (b) SPR curves of RIOs from 1.44 to 1.49 RIU to estimate the sensitivity and LOD. (c) Resonance angle comparison between simulation and experimental results.

The results achieved for the measurements of the RIOs short RI range set are depicted in Figure 9. In general, for all the liquids a small shift towards a higher resonance angle was observed of approximately 0.4±0.18°, which might be caused by the deviation of gold's dielectric constant [35]. Nevertheless, the results demonstrate good agreement with the intended results of the design. A difference of 7.5% was found in comparison to the simulated data. The sensitivity had a value of 171.87°/RIU and was calculated as the slope of the angles against the refractive index curve [53]. Furthermore, different standard deviation (σ) values were calculated for each RIO measured, where the minimum was 0.014° for RIO 1.4458 and the largest was 0.159° for RIO 1.49. A dependence was observed between the σ and the RIO which agrees with the findings of Canning [39]. Part of the error is due to the RI stability of the liquid, which might be particularly affected for certain RIOs. The LOD of the sensor was calculated with an average of the σ obtained for the lower RI 1.44 to 1.46, which was 0.028°. Then the value calculated for the LOD was $4.88 \times 10^{-4}$ RIU. In comparison to other SPR sensors ($10^{-4}$ to $10^{-7}$ RIU) [54], the LOD achieved might be high, nevertheless, further adaptations can be added to enhance this parameter if needed [55-59].

## 3.3 Polymer measurements

To measure the SPR curves of the selected polymers, the optical sensor was aligned to position the laser spot on the polymer sample's contact surface. The alignment was achieved by expanding the laser beam to ~1 cm and observing the resonance area with a CCD detector (Figure 10a). The resonance region was identified by shifting the polarization between P (for SPR) and S (to remove SPR) orientations to detect intensity changes (Figures 10a and 10b). Once identified, the beam was focused on the corresponding pixels of the detector (Figure 10c). The intensity signal was verified as SPR by toggling the polarizations again (Figures 10c and 10d). After alignment, the SPR curves of the polymers were measured similarly to the RIOs (see Figure 11). The obtained resonance angles are depicted (Table 2).

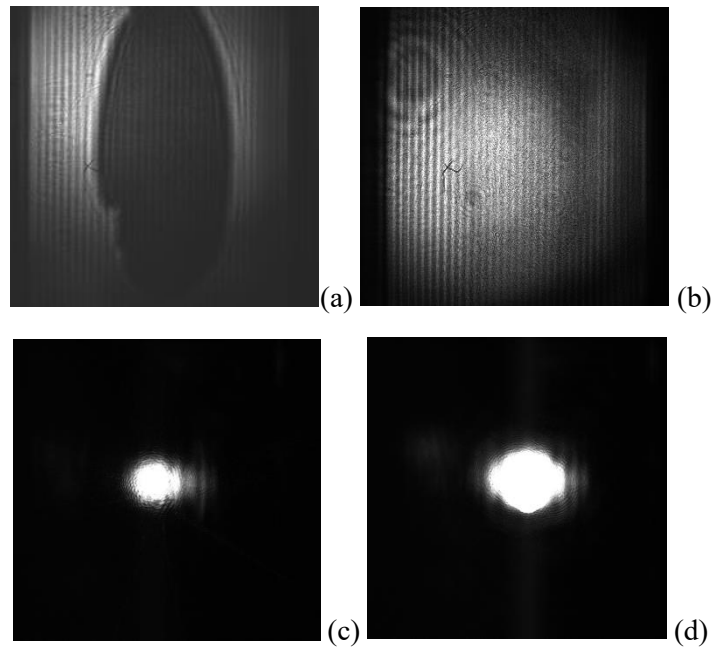

*Figure 10 Picture of a PMMA disk pressed against the sensor surface illuminated with an expanded beam in (a) P polarization and (b) S polarization. Next, the same region is illuminated with the laser spot in (c) P polarization and (d) S polarization.*

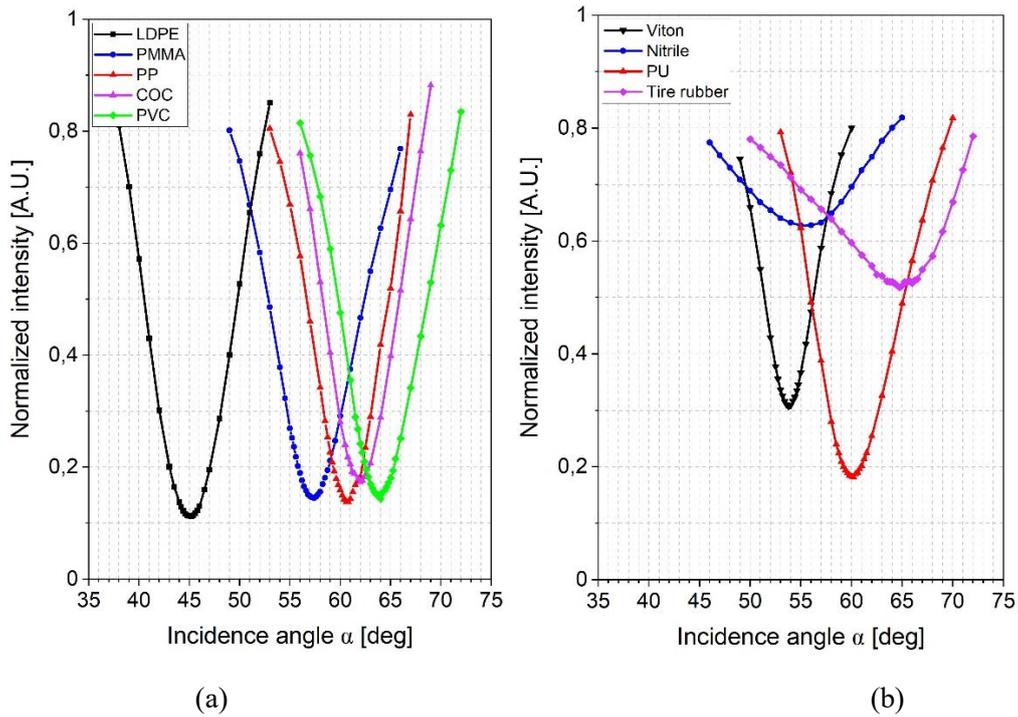

*Figure 11 SPR curves were measured for (a) polymer samples LDPE, PMMA, PP, COC, PVC and (b)Viton, Nitrile, PU, and Tyre rubber.*

*Table 2 Resonance angles were measured for different polymer types.*

| Polymer | Resonance angle ± σ*(°) | Polymer | Resonance angle ± σ (°) | Polymer | Resonance angle ± σ (°) |
|---|---|---|---|---|---|
| LDPE | 45,67 ± 0.01 | PMMA | 57,31 ± 0.01 | COC | 61,96 ± 0.02 |
| Viton | 53,91 ± 0.05 | PU | 60,16 ± 0.07 | PVC | 63,98 ± 0.05 |
| Nitrile | 55,00 ± 0.09 | PP | 60,70 ± 0.02 | TR | 65,06 ± 0.15 |

*σ Standard deviation

In general, the SPR curves acquired demonstrate the capability of the sensor designed to detect various polymers with SPR based on their dielectric constant, which is a novel result that to the best of our knowledge has not been reported before. In agreement with the RI measurements performed by [38], various polymers presented different resonance angles, which may lead to polymer identification with SPR. However, PP and PU presented similar resonance angles, therefore, further analysis should be done to support this claim. Discrepancies were evident among the minimum resonance intensity of the refractive index polymer curves. The lowest resonance minimum was obtained for the polymers that had better surface contact with the sensor. On the contrary, other materials such as TR, Viton and nitrile did not present an equally low resonance minimum intensity, which might also lead to a larger error in the resonance angle measured. A possible reason might be the surface waviness of the sample which may have prevented good contact between the sample and the sensor. Another explanation might be that the particles are made of a blend of polymers, which is the case of TR, therefore not all the surface area in contact presents resonance. Unfortunately, the SPR curve of PS could not be detected due to TIR in the sensor layers. As a final observation, it was noted that SPR is capable of analysing samples with black pigmentation, potentially providing a solution for analysing these materials.

A different error was expected for the polymer measurements in comparison to the RIOs measurements. The difference between these methods is that each polymer was measured in a different SPR chip and required the laser's alignment with the polymer's surface contact. Therefore, an additional experiment was made to calculate the repeatability of the sensor in these conditions. To assess the new error the resonance angle of the same PMMA sample was measured for four different SPR chips three times, aligning the laser spot each time. The results are depicted in Figure 12. From this analysis, the standard error associated with the laser alignment was 0.062°. On the other hand, the standard error associated with the SPR chip used was 2.971°, which demonstrates that the change of the SPR chip has a considerable impact on the measured resonance angle. In this specific experiment, σ equals 0.57°, which is an acceptable error in comparison to the angle range of polymer SPR resonance (45.67° to 65.06°). The curves collected with this method are expected to serve as a guide to enable the identification of the particle's polymers. However, microplastic samples can be analysed in a single alignment and one SPR chip, avoiding the error or the polymer curves collection.

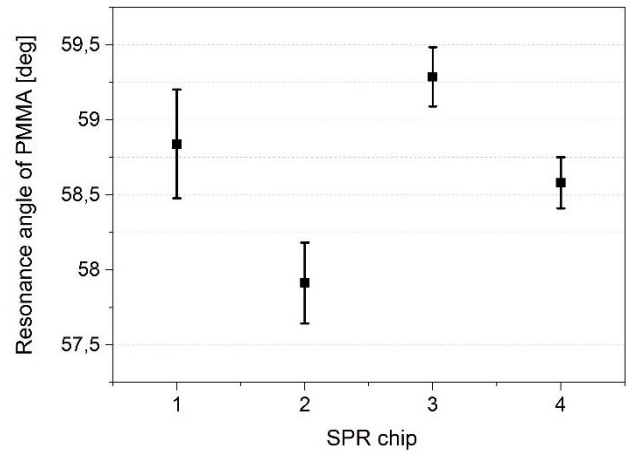

*Figure 12 Polymer measurements error due to SPR chip change and sample alignment.*

A particular advantage of the Kretschman configuration is its wide field of view (~1cm$^2$) [35] that enables the visualisation of the sample analysed. Based on the measurements conducted, the design demonstrated potential for visualizing the sample in contact with the sensor, as shown in Figure 10a. This capability was further explored by observing the SPR resonance image of a Viton and a PU sample (~5mm largest dimension) for various angles of incidence (see Figure 13). The frames illustrate the intensity of the polymer's SPR signal at a specific angle of the light beam incidence. The SPR resonance minimum intensity of Viton was observed at 54° and for PU at 62°, in agreement with the resonance values of Table 2. Moreover, these results depict how the SPR signal behaves equally for both materials regardless of their pigmentation and optical characteristics.

| Sample view | SPR frames in P polarization | | | |
|---|---|---|---|---|
| 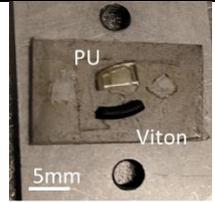 | 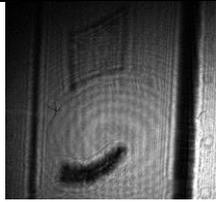 | 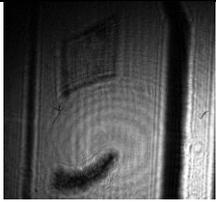 | 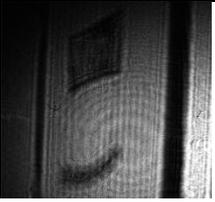 | 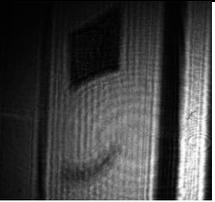 |
| Angles | 54° | 56° | 59° | 62° |

*Figure 13 Frames illustrating the shift in SPR resonance for pieces made of PU and Viton*

The frames in Figure 13 illustrate how the knowledge of the polymer resonance angles from the SPR curves can be combined with SPR imaging to acquire simultaneously more information about the particles for mixed polymer samples. These results also confirm that the SPR approach to microplastic analysis could be further enhanced to assess all the required parameters with the developments presented in this work. Further improvements in the resolution can be achieved through the selection of the output surface angle and the addition of an imaging system between the prism and the detector, as described by [60].

Aside from the improvement of the SPR approach for microplastic analysis, these findings may lead to other advantages such as a faster analysis time given that no spectra analysis is needed. In consequence, the potential benefits of using Surface Plasmon Resonance (SPR) for microplastic analysis warrant additional investigation, particularly to examine other possible advantages in detecting these contaminants. Further research is recommended to determine the viability of using SPR as an identification method for polymers. Additionally, there are still challenges to overcome to detect microparticles of TR with SPR such as finding a binding element to the sensor. Therefore, additional exploration should be made on the measurement of TR and similar materials with SPR.

## 4. Conclusions

This work demonstrated the capability of Surface Plasmon Resonance (SPR) to detect polymers using an angular interrogation approach. The dynamic range of the SPR sensor was examined through the optical design of a novel prism, which was based on the Kretschmann configuration. The effects of the input surface angle on the sensor's dynamic range were analysed. Experimental measurements showed that the sensor could measure refractive indices between 1.402 and 1.56 RIU. Notably, SPR could detect polymers regardless of their pigmentation, extending its use beyond traditionally transparent polymers.

Additionally, the prism enabled sample visualisation, which could, in the future, allow SPR to determine particle characteristics and facilitate microplastic analysis. To achieve this, it will be necessary to integrate the prism into an SPR imaging setup. This would enable the measurement of additional microplastic parameters, such as size, shape, and quantity, providing a more comprehensive evaluation of microplastics. These findings suggest a promising new approach to microplastics research. Further analysis could explore the effects of sample processing, polymer identification, and tyre rubber detection, ultimately enabling the SPR system to analyse real samples.

**Acknowledgement**

This work was part of the MSCA-ITN MONPLAS, which was supported by the European Union's Horizon 2020 research and innovation program under the Marie Skłodowska-Curie grant agreement No 860775. H2020-MSCA-ITN-2019.